\documentclass[reprint,superscriptaddress,amsmath,amssymb,
aps,prb,]{revtex4-2}
\usepackage[english]{babel}
\usepackage{amsmath}
\usepackage{amssymb}
\usepackage[bookmarks=true,colorlinks,citecolor=blue,linkcolor=blue,allcolors=blue]{hyperref}
\usepackage[capitalise]{cleveref}
\usepackage{physics}
\usepackage[toc,page]{appendix}
\usepackage{graphicx}
\usepackage[switch]{lineno}
\usepackage{orcidlink}
\usepackage{xurl}

\begin{document}


\title{Thermalization Dynamics in the Two-Dimensional Hubbard Model \\ with Neural-Network Quantum States}

\author{Alessandro Sinibaldi \orcidlink{0009-0009-8498-6068}}
\email{alessandro.sinibaldi@epfl.ch}
\affiliation{Institute of Physics, \'{E}cole Polytechnique F\'{e}d\'{e}rale de Lausanne (EPFL), CH-1015 Lausanne, Switzerland}
\affiliation{Center for Quantum Science and Engineering, EPFL, Lausanne, Switzerland}

\author{Luciano Loris Viteritti \orcidlink{0009-0004-2332-7943}}
\email{luciano.viteritti@epfl.ch}
\affiliation{Institute of Physics, \'{E}cole Polytechnique F\'{e}d\'{e}rale de Lausanne (EPFL), CH-1015 Lausanne, Switzerland}
\affiliation{Center for Quantum Science and Engineering, EPFL, Lausanne, Switzerland}

\author{Riccardo Rende \orcidlink{0000-0001-5656-4241}}
\affiliation{Center for Computational Quantum Physics, Flatiron Institute, 162 5th Avenue, New York, NY 10010}

\author{\\ Fakher F. Assaad \orcidlink{0000-0002-3302-9243}}
\affiliation{Institut f\"ur Theoretische Physik und Astrophysik, Universit\"at W\"urzburg, 97074 W\"urzburg, Germany}
\affiliation{W\"urzburg-Dresden Cluster of Excellence ctd.qmat, Am Hubland, 97074 W\"urzburg, Germany}

\author{Giuseppe Carleo \orcidlink{0000-0002-8887-4356}}
\affiliation{Institute of Physics, \'{E}cole Polytechnique F\'{e}d\'{e}rale de Lausanne (EPFL), CH-1015 Lausanne, Switzerland}
\affiliation{Center for Quantum Science and Engineering, EPFL, Lausanne, Switzerland}

\date{\today}

\begin{abstract}
Thermalization in strongly correlated fermionic systems remains a central open problem in quantum many-body physics.
In this work, we investigate the real-time dynamics and the approach to thermalization in the two-dimensional Hubbard model, a paradigmatic framework for correlated electrons, relevant to high-temperature superconductivity and ultracold quantum simulation.
Focusing on the half-filled square lattice, we monitor the time evolution of the double occupancy following a quench in the on-site interaction $U$, and assess whether its long-time value is captured by a canonical thermal ensemble.
We employ time-dependent variational Monte Carlo methods combined with transformer-based Neural-Network Quantum States to accurately describe the nonequilibrium dynamics of fermions, especially for the behavior at long times, thereby accessing regimes that were previously inaccessible to numerical simulations.
Our results reveal two dynamical behaviors: for weak to intermediate interactions, the double occupancy rapidly approaches the thermal prediction, consistent with ergodic evolution; beyond a critical interaction $U_{C}$, the dynamics remains distinct from the thermal expectation on the timescales investigated, revealing signatures of a prethermal plateau delaying fast relaxation.
These results establish numerical simulation as a powerful tool to probe nonequilibrium quantum phenomena in correlated fermionic matter.
\end{abstract}

\maketitle
\section{Introduction}
Understanding the real-time dynamics of interacting fermions is a central challenge in quantum matter.
Many of the key experimental questions in strongly correlated materials are now dynamical~\cite{zong2023emerging}.
Time-resolved measurements can show how electronic order is destroyed, how spectral weight is redistributed, and how charge, spin, and lattice degrees of freedom exchange energy after a perturbation~\cite{giannetti2016ultrafast,delatorre,basov}.
In the cuprates, for example, ultrafast experiments have probed the dynamics of Cooper pairs and the recovery of the superconducting gap after a laser pulse~\cite{smallwood2012tracking}, while nonlinear optical studies have reported light-enhanced coherent transport~\cite{hu2014optically}. 
Related experiments have directly observed ultrafast photoinduced insulator-to-metal transitions in Mott materials~\cite{amano2024propagation,verma2024picosecond}.
These advances have sharpened a basic question: \textit{which aspects of nonequilibrium behavior are specific to a given material, and which are intrinsic to strongly interacting fermions?}
Answering this question in real materials is challenging because of the inevitable presence of many competing effects, including disorder, phonons, multiband structure, and coupling to external environments.

\begin{figure*}[t]
    \centering \includegraphics[width=1.0\linewidth]{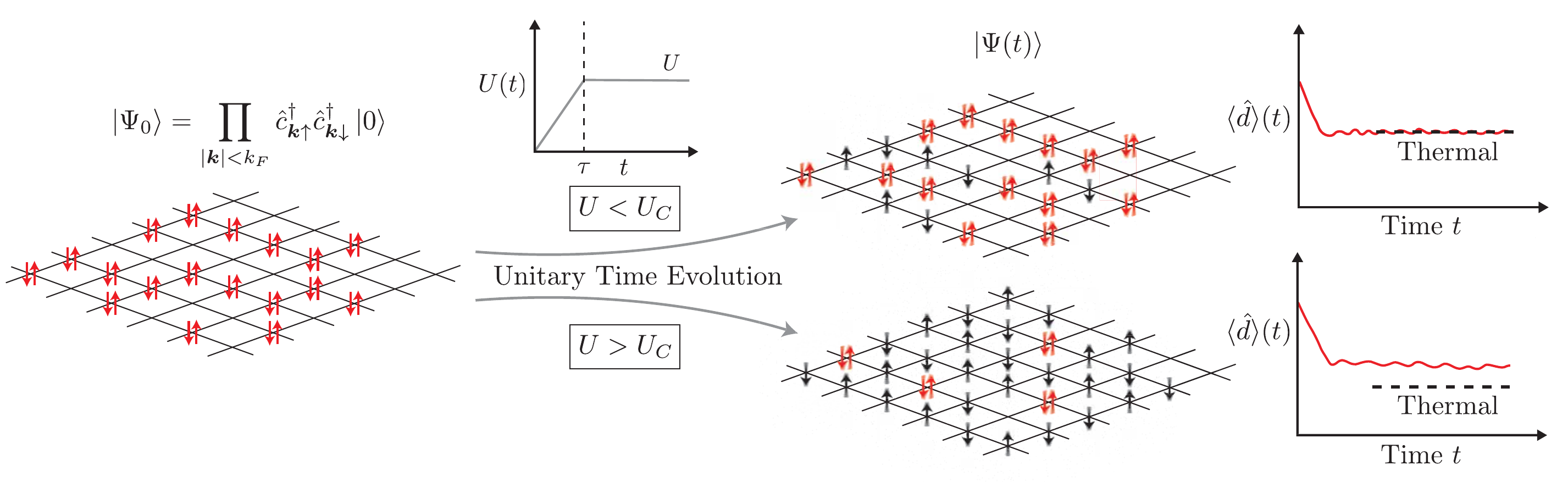}
    \caption{Schematic illustration of the dynamical protocol studied to investigate thermalization. 
    An initial Fermi sea state is evolved using the Hubbard Hamiltonian in~\cref{eq:hamiltonian} with the interaction strength $U(t)$ following a linear ramp quench with final interaction $U$.
    We expect that for $U$ smaller than a critical value $U_C$ the double occupancy $\langle{\hat{d}\rangle}(t)$ [see \cref{eq:double_occ}] at long time relaxes to the value predicted by thermalization, whereas for $U$ larger than $U_C$ it departs from the thermal prediction.}
    \label{fig:artistic}
\end{figure*}

Quantum simulators based on ultracold atoms provide a complementary route to the study of strongly correlated fermions.
They realize clean, isolated, and highly tunable systems in which geometry, density, interaction strength, and driving protocols can be controlled with a level of precision that is difficult to achieve in real materials~\cite{schneider2012fermionic,daley2022practical}.
Over the past decade, these platforms have evolved from proof-of-principle emulators into precision probes of correlated quantum matter.
Experiments have realized both Mott-insulating and metallic regimes~\cite{jordens2008mott,schneider2008metallic}, directly measured spin and charge correlations with single-site resolution~\cite{parsons2016site,cheuk2016observation,boll2016spin}, and, more recently, accessed nonequilibrium transport~\cite{nichols2019spin,schneider2012fermionic,brown2019bad}, magnetic polarons~\cite{koepsell2019imaging}, and long-range fermion pairing~\cite{hartke2023direct}.

A further route, complementary to both real-material experiments and quantum simulators, is provided by classical numerical simulations.
These are essential both for interpreting experimental observations and for proposing new setups that can subsequently be tested experimentally.
More importantly, numerical calculations, much like quantum simulators, provide access to simplified settings in which specific instances of nonequilibrium dynamics can be isolated and studied in a controlled way.
Despite these advantages, the numerical study of real-time quantum dynamics is notoriously difficult in interacting fermionic systems~\cite{daley2022practical}.
In the two limiting cases of infinite dimensions and one dimension, Dynamical Mean-Field Theory (DMFT)~\cite{georges1992hubbard} and Tensor Networks  (TN)~\cite{white1992density}, respectively, yield accurate results.
In two dimensions, however, which is both directly relevant to quantum simulators and hosts some of the richest correlated phenomena, no method currently captures the real-time dynamics of interacting fermions accurately beyond the small clusters accessible to exact diagonalization~\cite{dagotto,innerberger2020electron}.
Existing studies in two dimensions have been limited to free fermionic systems~\cite{wu2023tensor} and effective low-energy models~\cite{hubig2020evaluation} with TN-approaches, or to single-particle dynamics with a Heisenberg-picture algorithm restricted to evolutions with low-weight operators~\cite{d2025majorana}.

In this work, we investigate the many-body real-time dynamics of the two-dimensional Hubbard model, the most paradigmatic model of strongly interacting fermions, which provides a minimal framework for doped Mott physics and captures the key aspects of the phenomenology of high-$T_c$ superconductivity~\cite{arovas2022hubbard}.
Specifically, we focus on the investigation of thermalization, one of the central questions in the dynamics of strongly correlated quantum systems.
At the most basic level, thermalization refers to the relaxation of observables towards their thermal values after unitary time evolution, under ergodic hypotheses~\cite{srednicki,rigol2008thermalization}.
Its experimental relevance is immediate: it determines how the long-time state reached after a quench, pulse or ramp should be interpreted.
Therefore, it decides whether late-time measurements can be described in terms of an effective temperature and equilibrium thermodynamics, or instead reflect long-lived prethermal or genuinely non-thermal behaviour.
This, in turn, sets the timescales over which ordered states or metastable regimes can persist and directly shapes the interpretation of transport measurements, spectroscopic observables, and state-preparation protocols~\cite{kaufman2016quantum,reimann2016typical,RevModPhys.91.021001}. 

Earlier studies on the Hubbard model in infinite dimensions, based on nonequilibrium DMFT, have shown that thermalization can be delayed by prethermal plateaux, strongly modified near dynamical critical points, or even prevented by trapping in long-lived non-thermal ordered states~\cite{werner1,werner2,werner3,werner4,werner5,murakami2025photoinduced}.
Understanding which, if any, of these scenarios is realized in two dimensions, precisely where the physics is most relevant to cuprates and most accessible to quantum simulators, remains an outstanding open question.

Here, we perform numerical simulations of ramp-quench dynamics of the Hubbard model at half filling, and then compare the long-time extrapolations of the double occupancy, a relevant local observable, with numerically exact finite-temperature Quantum Monte Carlo~\cite{blankenbecler1981monte} results.
To enable accurate and scalable simulations of far-from-equilibrium many-body dynamics, especially in the long-time regime, we combine time-dependent Variational Monte Carlo~\cite{carleo_localization_2012, becca2017} with the time-dependent Linear Variational Method~\cite{motta2024subspace,sinibaldi2024time}, using a powerful Neural-Network Quantum State ansatz~\cite{carleo2017solving} based on the transformer architecture~\cite{viteritti2023transformer_1d}.
Previous works have performed variational calculations on the Hubbard model for the phenomenon of laser-driven superconductivity~\cite{ido2015time,ido2017correlation}, but those studies relied on physically motivated but less expressive ans\"{a}tze and did not address thermalization.
The combination of a highly expressive neural Ansatz with a robust time-evolution scheme enables accurate simulations at the system sizes and timescales required to investigate thermalization, extending the capabilities of numerical simulations to regimes inaccessible to current computational methods. 

Our main finding is the existence of a dynamical regime in which the extrapolated long-time values differ from their thermal counterparts, providing the first numerical signature of a delay of thermalization in the two-dimensional Hubbard model, compatible with a prethermal plateau.

\begin{figure*}[hbtp]
    \centering \includegraphics[width=1.0\linewidth]{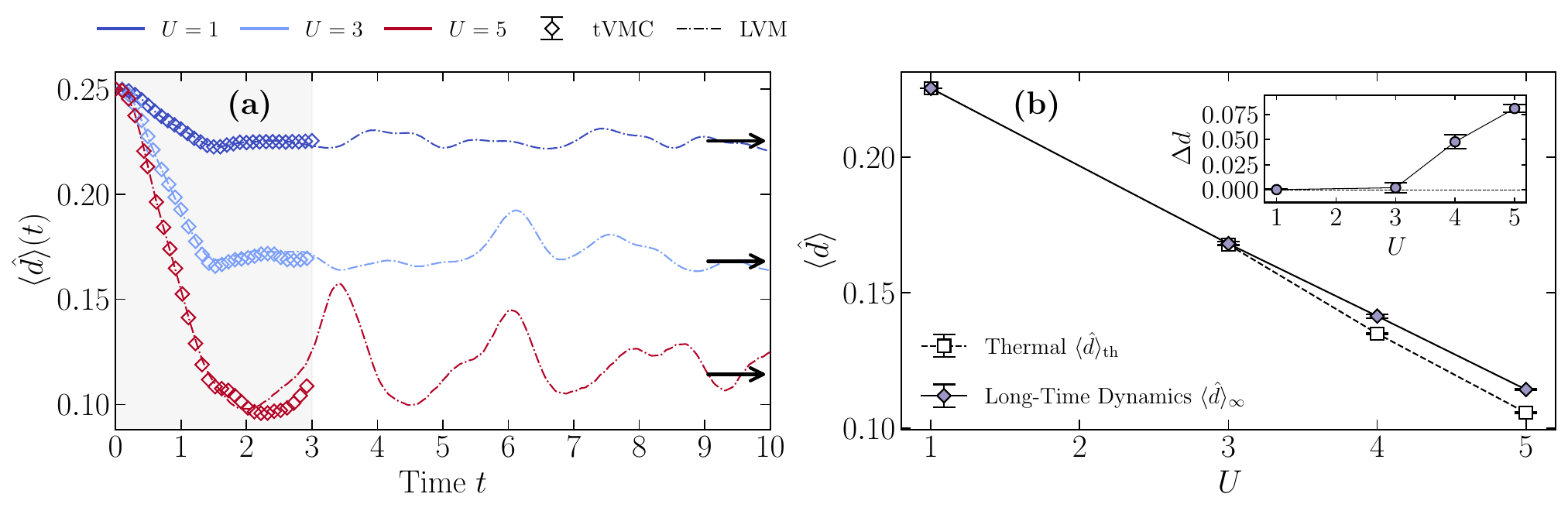}
    \caption{\textbf{Panel (a)}: Time evolution of the double occupancy $\langle \hat{d} \rangle (t)$ [see \cref{eq:double_occ}] for various interaction strengths $U$ on a $8 \times 8$ lattice.
    The empty markers denote the tVMC simulations, while dashed lines indicate the dynamics extended via the time-dependent LVM. The shaded region highlights the time interval from which tVMC states are selected to perform the LVM computation. 
    The arrows show the infinite-time limit of the observable from LVM.
    The Monte Carlo errors are smaller than the markers and are not shown for clarity. 
    \textbf{Panel (b)}: Infinite-time value of the double occupancy $\langle \hat{d} \rangle_{\infty}$ compared with the effective thermal value $\langle \hat{d} \rangle_{\text{th}}$ as a function of $U$. 
    The inset displays the order parameter $\Delta d$ [see \cref{eq:order_par}].
    The statistical error on $\langle \hat{d} \rangle_{\infty}$ is estimated by repeating the LVM calculation over independent sets of Monte Carlo samples.
    }
    \label{fig:therm}
\end{figure*}

\section{Non-equilibrium protocol}
In this section, we introduce the physical model and the nonequilibrium protocol considered in this work.
The system consists of $N_e$ electrons on a two-dimensional ${L \times L}$ square lattice and $N=L^2$ sites.
Throughout, we focus on the sector at half-filling, namely $N_e=N$, and equal numbers of spin-up and spin-down particles (zero total spin-$z$ component).
The system is described by the time-dependent Hubbard Hamiltonian
\begin{equation}
\label{eq:hamiltonian}
\hat{H}(t) = -t_{\text{hop}} \sum_{\langle i,j \rangle,\sigma} 
\left(\hat{c}^{\dagger}_{i\sigma} \hat{c}_{j\sigma} + \mathrm{h.c.} \right)
+ U(t) \sum_{i} \hat{n}_{i\uparrow} \hat{n}_{i\downarrow},
\end{equation}
where $\hat{c}^{\dagger}_{i\sigma}$ ($\hat{c}_{i\sigma}$) creates (annihilates) an electron with spin $\sigma \in \{\uparrow,\downarrow\}$ on site $i$, and $\hat{n}_{i\sigma} = \hat{c}^{\dagger}_{i\sigma} \hat{c}_{i\sigma}$ is the corresponding number operator.
Here $t_{\text{hop}}$ is the nearest-neighbor hopping amplitude and $U(t)$ denotes the time-dependent on-site interaction. In what follows, we set the hopping amplitude to $t_{\text{hop}} = 1$ and express all remaining energy scales in units of $t_{\text{hop}}$.

We study the real-time dynamics obtained by preparing at $t=0$ the system in the non-interacting ground state $\ket{\Psi_0}$, that is, the ground state of Eq.~\eqref{eq:hamiltonian} at ${U(t=0)=0}$, in periodic or antiperiodic boundary conditions according to the system size so that the corresponding non-interacting system satisfies the closed-shell condition. Then, we drive it out of equilibrium through a linear ramp of the interaction up to a final value $U$ over a finite ramp time $\tau$.
The time-dependence of the interaction strength is therefore
\begin{equation}
\label{eq:protocol}
U(t) =
\begin{cases}
U \dfrac{t}{\tau}, & 0 \le t \le \tau, \\[6pt]
U, & t > \tau.
\end{cases}
\end{equation}

The protocol~\cref{eq:protocol} interpolates between two standard nonequilibrium settings: in the limit $\tau \to 0$, the ramp becomes a sudden quench to the Hubbard Hamiltonian with interaction strength $U$, while for $\tau \to \infty$ it approaches an adiabatic preparation of the interacting state. Our purpose here is not to perform a systematic analysis of different driving protocols, but to consider a fixed realistic protocol to test whether and under what conditions fast thermalization breaks down. We therefore adopt, following Ref.~\cite{werner3}, the linear ramp in \cref{eq:protocol} and set $\tau = 1.25$ in all calculations.

\section{Thermalization of local observable} 
We analyze the dynamics of a local observable evolving with the protocol in~\cref{eq:protocol} and the approach to thermalization at long times as a function of the final interaction strength $U$.
The local observable we consider is the double occupancy
\begin{equation}
\label{eq:double_occ}
    \hat{d} = \frac{1}{N} \sum_{i} \hat{n}_{i\uparrow}\hat{n}_{i \downarrow} \ .
\end{equation}

Previous results on the time evolution of quantum many-body systems~\cite{carleo_localization_2012,sinibaldi2024time,blas_test_2016,werner2,werner3,werner4,murakami2025photoinduced} suggest that strong quench dynamics can bring the initial state to long-lived metastable states where observables differ from their thermal expectations, leading to an effective delay of thermalization.
A schematic illustration of our dynamical protocol and the expected thermalization behaviour is provided in \cref{fig:artistic}.

In this work, the time evolution in the two-dimensional Hubbard model is simulated using the time-dependent Variational Monte Carlo (tVMC) method~\cite{carleo2012localization} combined with a Neural-Network Quantum State (NQS) wave function~\cite{carleo2017solving} based on the backflow transformer architecture~\cite{gu2025, viteritti2026beyond}. 
To access the long-time limit of the dynamics and probe thermalization, we employ the procedure introduced in Ref.~\cite{sinibaldi2024time} based on the time-dependent Linear Variational Method (LVM).
This approach solves the Schrödinger equation in a fixed set of basis states and drops fast oscillating terms in the coefficients of the superposition. 
The basis states that we consider are states evolved at different times through tVMC. 
In particular, we use states that are equally spaced in time, such that they approximate a Krylov basis generated by repeated application 
of the time-step propagator, ${\ket{\Phi_k}  = (e^{-i \Delta t \hat{H}})^k \ket{\Psi_0}}$.
The number of basis states is a key control parameter governing the accuracy of the method, and extrapolations in this parameter can be performed to fully converge the results.
Details on the tVMC approach and the transformer NQS architecture are provided in \cref{app:finite_time}, whereas a complete description of the LVM is provided in~\cref{app:infinite_time} of the \textit{Methods}.

The extracted long-time expectation values are compared with the thermal averages computed using Auxiliary-Field Quantum Monte Carlo (AFQMC)~\cite{blankenbecler1981monte} 
in the canonical ensemble~\cite{wang2017finite}. 
The effective temperature $T_{\mathrm{eff}}$ is determined by requiring the thermal energy to match the energy reached by the system after the ramp quench. 
This corresponds to solving for $T_{\text{eff}}$ in the equation
\begin{equation}
\label{eq:t_eff}
    \left. \frac{\text{Tr}[e^{\hat{H}(t) / T_{\text{eff}}} \hat{H}(t)]}{\text{Tr}[e^{\hat{H}(t) / T_{\text{eff}}}]} \right|_{t = \tau} = \left. \frac{\bra{\Psi(t)} \hat{H} (t) \ket{\Psi(t)}}{\bra{\Psi(t)} \ket{\Psi(t)}} \right|_{t = \tau}, 
\end{equation}
where $\ket{\Psi(t)}$ is the time-evolved state at time $t$.
The thermal double occupancy $\langle \hat{d} \rangle_{\text{th}}$ is obtained as the thermal expectation value of the operator $\hat{d}$ at this temperature $T_{\text{eff}}$, namely ${\langle \hat{d} \rangle_{\text{th}} = \text{Tr}[e^{\hat{H}(t) / T_{\text{eff}}}\hat{d}] / \text{Tr}[e^{\hat{H}(t) / T_{\text{eff}}}]}$ for $t=\tau$.
The finite-temperature calculations are described in more detail in~\cref{app:finite_temperature} of the \textit{Methods}.

The main results of our work are presented in~\cref{fig:therm}.
The panel (a) shows the tVMC dynamics (empty markers) of the double occupancy $\langle \hat{d} \rangle (t)$ for various interaction strengths $U$ on the $8 \times 8$ lattice.
States computed up to $t=3$ in units of $t_{\text{hop}}$ (highlighted region) are employed as the basis for the LVM calculation (dashed lines).
The infinite-time value $\langle \hat{d} \rangle_\infty = \lim_{t \rightarrow \infty} \langle \hat{d} \rangle (t)$ extracted from the LVM, indicated by an arrow, is coherent with the average of the trajectory at finite time.
Additionally, we observe that as $U$ increases, $\langle \hat{d} \rangle (t)$ reaches lower values. 
This is expected by considering that, in the limit when the protocol~\cref{eq:protocol} is adiabatic, namely for $\tau \gg 1$, the time evolved state at each time $t$ coincides with the ground state of~\cref{eq:hamiltonian} with $U(t)$.
Therefore, since larger $U$ penalizes double occupancy in the ground state, $\langle \hat{d} \rangle (t)$ is expected to decrease with the interaction strength. 
In the non-adiabatic regime relevant here, we expect the system to end up in a superposition of eigenstates with lower double occupancy as $U$ is greater.

~\cref{fig:therm}(b) compares $\langle \hat{d} \rangle_{\infty}$ with the effective thermal average $\langle \hat{d} \rangle_{\text{th}}$ for different values of $U$.
For $U \leq 3$, the two quantities agree closely, indicating that the dynamics is ergodic and the system thermalizes at long times.
For $U > 3$, however, the dynamical values deviate appreciably from the thermal expectations, suggesting a markedly different thermalization behavior in this regime.

Based on the discussion above, a suitable order parameter for identifying the transition is 
\begin{equation}
\label{eq:order_par}
\Delta d = \left|1 - \frac{\langle \hat{d} \rangle_{\infty}}{\langle \hat{d} \rangle_{\text{th}}}\right| \ , 
\end{equation}
which vanishes when thermalization occurs and becomes finite when thermalization does not strictly apply.
This quantity is shown in the inset of~\cref{fig:therm}(b), highlighting that the nonthermal behavior becomes increasingly pronounced at stronger interactions.
The critical interaction strength $U_{C}$ above which fast thermalization no longer occurs is estimated to lie between $U=3$ and $U=4$.
The observed phenomenology beyond $U_{C}$ is compatible with a long-lived prethermal plateau that delays thermalization on the time scales investigated.
This slow relaxation is consistent with the exponentially long lifetime of double occupancies at large interactions previously predicted theoretically~\cite{eckstein,rosch2008metastable} and observed experimentally~\cite{sensarma2010lifetime}.
From our results, we cannot claim a complete breakdown of thermalization, since the infinite-time values are extrapolated from finite-time simulations, which by construction cannot resolve the dynamics at arbitrarily late times.
We remark that this caveat is inherent to any numerical investigation of thermalization based on time evolution: eventual convergence to the thermal value at very late times can never be fully excluded.
In any case, the behavior we observe above $U_{C}$ is clearly distinct from the fast thermalization occurring for $U<U_{C}$, suggesting the onset of a nontrivial dynamical regime beyond the critical interaction strength.

\begin{figure}[t]
    \centering \includegraphics[width=1.0\linewidth]{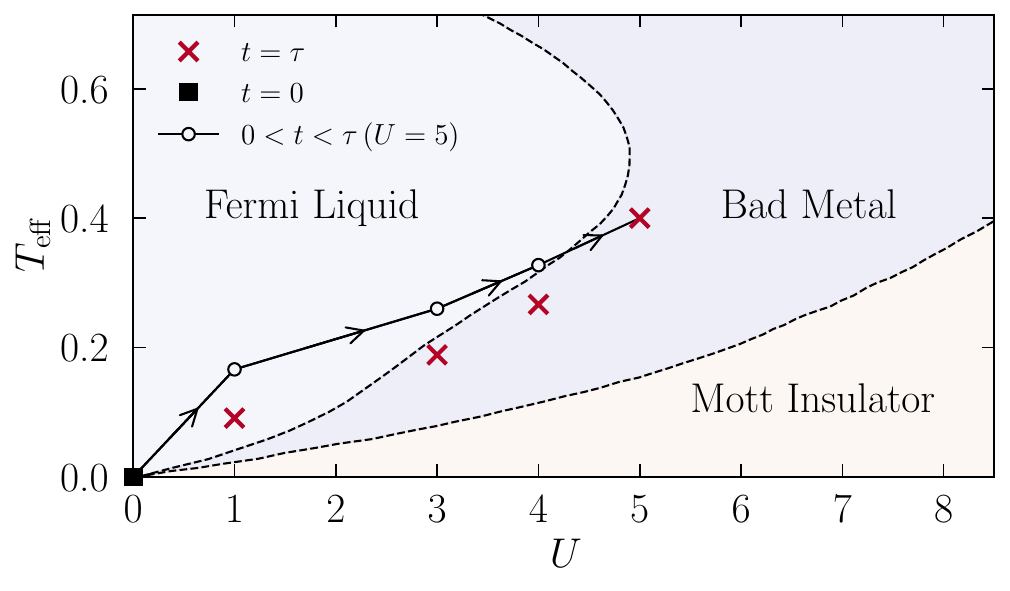}
    \caption{Effective temperature $T_{\mathrm{eff}}$ as a function of the interaction strength $U$ for the dynamics with $U=1, 3, 4, 5$ shown in~\cref{fig:therm} at times $t=0$ (black square) and $t=\tau$ (red crosses).
    The dashed lines indicate the boundaries of the finite-temperature putative phase diagram of Ref.~\cite{lu2026quantum}.
    Points at intermediate times (empty circles) are also shown for $U = 5$ to illustrate the trajectory through the phase diagram during the dynamics.
    The arrows indicate the direction of time. 
    }
    \label{fig:phase_diagram}
\end{figure}

To better interpret our results, it is useful to locate the effective temperatures reached by the dynamics within the known finite-temperature phase diagram of the Hubbard model at half filling.
\cref{fig:phase_diagram} shows the values of $T_{\mathrm{eff}}$ corresponding to the dynamics of \cref{fig:therm} at different interaction strengths, superimposed on the putative crossover phase diagram computed with AFQMC in Ref.~\cite{lu2026quantum} .
The point at $t=0$ is located at the origin, since the initial state is the ground state of $\hat{H}(t=0)$.
We observe that, as $U$ increases, the final effective temperature crosses from the Fermi-liquid regime into the bad-metal phase. 
This seems to suggest that the thermalization phenomenology observed at large $U$ is connected to the qualitatively different nature of the phase reached by the dynamics.

Similar phenomenology to that found here has been reported in interacting spin systems~\cite{blas_test_2016,sinibaldi2024time} and lattice bosons~\cite{carleo_localization_2012}, but it had not previously been observed in two-dimensional correlated fermionic systems.
Our findings extend previous DMFT results ~\cite{werner1,werner2,werner3,werner4,werner5,murakami2025photoinduced}, which report the same physics in infinite dimensions.
Our work demonstrates that the prethermalization plateau in the dynamics of the Hubbard model is a robust physical phenomenon that persists in dimensionality directly relevant to real experimental setups.

\section{Conclusions}

In this work, we have investigated the real-time dynamics of the two-dimensional Hubbard model after an interaction quench and the approach to thermalization at long-times.
By comparing extrapolated infinite-time values of a local observable with finite-temperature Quantum Monte Carlo results, we find evidence for a long-lived delay of thermalization above a critical interaction strength $U_{C}$ lying between $U=3$ and $U=4$.
For $U \leq U_{C}$, the dynamics converges to the thermodynamic prediction, whereas for larger $U$ the long-time values detach markedly from the thermal expectations on the accessible timescales.
These findings extend earlier nonequilibrium DMFT studies~\cite{werner1,werner2,werner3,werner4,werner5,murakami2025photoinduced}, which identified analogous phenomenology in infinite dimensions.
Our results demonstrate that the absence of fast thermalization at strong interactions persists robustly in two dimensions, where the interplay of strong correlations and finite connectivity is most relevant to realistic cuprate physics~\cite{hu2014optically,smallwood2012tracking,amano2024propagation,verma2024picosecond}.
At the same time, our work is especially relevant given that ultracold fermionic atoms in optical lattices now offer pristine realizations of the two-dimensional Hubbard model down to cryogenic temperatures, making the non-equilibrium dynamics of this model directly accessible to experiments~\cite{schneider2012fermionic,daley2022practical,jordens2008mott,schneider2008metallic,parsons2016site,cheuk2016observation,boll2016spin,nichols2019spin,brown2019bad,koepsell2019imaging,hartke2023direct}.

From a methodological standpoint, this work establishes the combination of time-dependent variational methods with transformer-based Neural-Network Quantum States as a viable framework for studying far-from-equilibrium fermionic dynamics in two dimensions, a regime where controlled numerical approaches have so far been lacking.
The ability to systematically extend the accessible time window in the dynamics positions this approach as a powerful tool for investigating the dynamical behavior of strongly correlated materials~\cite{zong2023emerging} and serves as a way to classically benchmark quantum simulators at long times~\cite{king2025, mauron2025, mauron2025predicting, tindall2026}.

Several directions remain open.
Extending the analysis to more realistic models that better capture cuprate physics is a natural next step. This includes studying doped systems and incorporating longer-range hopping amplitudes and additional interaction terms, which can generate a richer interplay between charge dynamics and magnetic correlations~\cite{white1989, gull2015, xu2024, gu2025, simkovic2024, Roth2025SCNQS, viteritti2026beyond}.
Finally, investigating spectral functions of correlated fermions~\cite{imadaspectral,feiguinspectral,rigo2025operator,PhysRevB.44.10256} from the real-time dynamics represents a fruitful next step.

\section*{Methods}

\subsection{Finite-Time Dynamics}
\label{app:finite_time}

\subsubsection{Time-Dependent Variational Monte Carlo}
\label{app:tdvp_tvmc}

The time evolution of a quantum state $\ket{\Psi(t)}$ governed by the Hamiltonian $\hat{H}$ is described by the time-dependent Schrödinger equation
\begin{equation}
\frac{d}{dt}\ket{\Psi(t)} = -i \hat{H} \ket{\Psi(t)} .
\label{eq:tdse}
\end{equation}

In variational approaches, the exact state $\ket{\Psi(t)}$ is approximated by a parameterized Ansatz $\ket*{\Psi_{\theta(t)}}$, whose time dependence is encoded in the variational parameters $\theta(t)$. Time-dependent variational principles (TDVP)~\cite{Yuan2019theoryofvariational} determine parameter updates such that $\ket{\Psi_{\theta(t)}}$ follows the dynamics dictated by \cref{eq:tdse}.
In McLachlan's formulation, this is achieved by minimizing the Euclidean distance between the two sides of~\cref{eq:tdse}, which yields the following linear system for the parameter velocities:
\begin{align}
\sum_{\beta} \Re[S_{\alpha\beta}]\,\dot{\theta}_{\beta} = \Re[F_\alpha].
\label{eq:tdvp}
\end{align}
Without loss of generality, we assume the parameters to be real~\footnote{The case of complex parameters can be handled by treating each complex parameter as a pair of real-valued parameters.}.
Here, $S$ denotes the quantum geometric tensor
\begin{equation}
\label{eq:s}
    S_{\alpha\beta} 
= \frac{\bra{\partial_{\alpha}\Psi_{\theta}}\ket{\partial_{{\beta}}\Psi_{\theta}}}{\braket{\Psi_{\theta}}{\Psi_{\theta}}}
- \frac{\bra*{\partial_{\alpha}\Psi_{\theta}}\ket{\Psi_{\theta}}}{\braket{\Psi_{\theta}}{\Psi_{\theta}}}
  \frac{\bra{\Psi_{\theta}}\ket{\partial_{\beta}\Psi_{\theta}}}{\braket{\Psi_{\theta}}{\Psi_{\theta}}},
\end{equation}
while $F$ is related to the energy gradient~\cite{becca2017}
\begin{equation}
\label{eq:f}
    F_\alpha 
= -i\left(
\frac{\bra{\partial_{\alpha}\Psi_{\theta}}\hat{H}\ket{\Psi_{\theta}}}{\braket{\Psi_{\theta}}{\Psi_{\theta}}}
- \frac{\bra{\partial_{\alpha}\Psi_{\theta}}\ket{\Psi_{\theta}}}{\braket{\Psi_{\theta}}{\Psi_{\theta}}}
  \frac{\bra{\Psi_{\theta}}\hat{H}\ket{\Psi_{\theta}}}{\braket{\Psi_{\theta}}{\Psi_{\theta}}}
\right) \ .
\end{equation}

For notational compactness, the explicit time dependence of the variational parameters has been omitted throughout.

In the case of correlated wave functions, both quantities~\cref{eq:s,eq:f} can be estimated efficiently by Monte Carlo sampling from the normalized Born probability distribution ${\Pi(\boldsymbol{n}) \propto |\Psi_\theta(\boldsymbol{n})|^2}$, where $\boldsymbol{n}$ denotes a fermion configuration in the occupation number basis~\cite{becca2017}.
This leads to the following stochastic expressions~\cite{rende2024simple}:
\begin{equation}
\label{eq:stoch_est}
    \text{Re}[S] = X X^T, \qquad \text{Re}[F] = X g,
\end{equation}
where ${X = \text{Concat}(\text{Re}[Y], \text{Im}[Y]) \in \mathbb{R}^{P \times 2M}}$ with ${Y_{\alpha i} = (O_{\alpha i} - \bar{O}_{\alpha}) / \sqrt{M}}$ and ${O_{\alpha i} = \partial_{\theta_{\alpha}} \log \Psi_{\theta}(\boldsymbol{n}_i)}$ is the Jacobian of the logarithm of the variational wave function.
Equivalently, ${g = \text{Concat}(\text{Im}[\varepsilon], -\text{Re}[\varepsilon]) \in \mathbb{R}^{2M}}$ with ${\varepsilon_i = (E_{\text{loc}, i} - \bar{E}_{\text{loc}}) / \sqrt{M}}$, where we have defined the local energy as ${E_{\text{loc}, i} = \bra{\boldsymbol{n}_i} \hat{H} \ket{\Psi_{\theta}} / \braket{\boldsymbol{n}_i}{\Psi_{\theta}}}$.
In the previous expressions, $\boldsymbol{n}_i$ denotes the $i$-th of $M$ Monte Carlo samples drawn from $\Pi(\boldsymbol{n})$, the bar denotes the empirical average over these samples, and $P$ is the number of variational parameters.
We note that $g$ differs from the analogous quantity used in Ref.~\cite{rende2024simple} for ground-state optimization, as here we consider the equations of real-time dynamics [see \cref{eq:tdvp}].

\begin{figure}[t]
    \centering \includegraphics[width=1.0\linewidth]{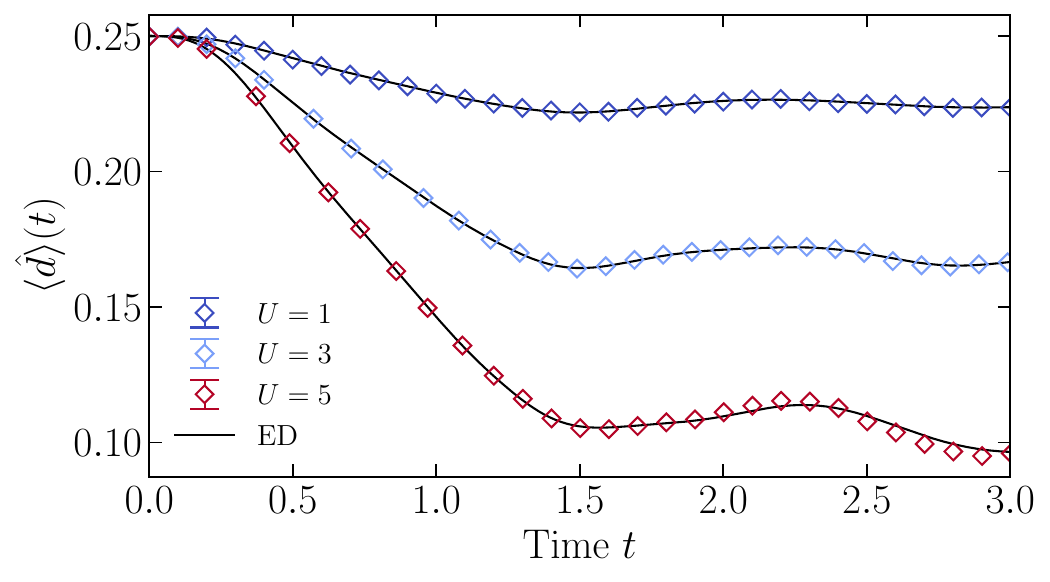}
    \caption{Time evolution with tVMC (empty markers) of the double occupancy $\langle \hat{d} \rangle(t)$ with different $U$ for the $4 \times 4$ lattice.
    The solid black lines represent the exact diagonalization (ED) results.
    The Monte Carlo errors are smaller than the markers and are not shown for clarity.} 
    \label{fig:4x4_benchmark}
\end{figure}

Since the matrix $\text{Re}[S]$ in~\cref{eq:stoch_est} has dimension $P \times P$, directly solving~\cref{eq:tdvp} by inverting it becomes computationally prohibitive in the deep learning regime~\cite{rende2024simple}.
Following the approach originally introduced for ground-state calculations~\cite{chen2024empowering,rende2024simple}, rather than solving ${\dot{\theta} = (X X^T)^{-1} Xg}$ one can solve the equivalent system
\begin{equation}
\label{eq:minsr}
    \dot{\theta} = X (X^T X)^{-1} g \ ,
\end{equation}
which requires inverting the $2M \times 2M$ matrix $X^T X$, known as the \emph{Neural Tangent Kernel} (NTK)~\cite{novak2022fast}.
Since $M \ll P$ in the deep learning regime, this reformulation renders the problem computationally tractable, enabling the use of variational wave functions with hundreds of thousands or even millions of parameters, as employed in this work.
In practice, $X^T X$ is symmetric and positive semi-definite, and is inverted by diagonalizing it and discarding eigenvalues smaller than $r_{\text{cond}} \cdot \lambda_{\text{max}}$~\cite{carleo_localization_2012,medvidovic2023variational,schmitt2020quantum}, where $\lambda_{\text{max}}$ is the largest eigenvalue and $r_{\text{cond}} = 10^{-9}$.
This regularization ensures numerical stability in the dynamics, as $X^T X$ can be ill-conditioned~\cite{becca2017}.

The parameter velocities $\dot{\theta}$ obtained from~\cref{eq:minsr} are used to update the variational parameters via a standard ordinary differential equation (ODE) integrator~\cite{schmitt2020quantum}.
The time step in the integration is chosen as $\delta t = 10^{-2}/U(t)$ when $U(t) > t_{\mathrm{hop}}$, and as $\delta t = 10^{-2}/t_{\mathrm{hop}}$ otherwise.
This ensures that the dominant term in the exponent of the unitary propagator remains constant throughout the time evolution~\cite{ido2015time}.
The impact of the choice of ODE integrator is analyzed in~\cref{app:hyperparams}.

We remark that all the previous expressions can be extended to time-dependent Hamiltonians $\hat{H}(t)$, as in the case considered in this work.
This requires approximating the Hamiltonian over each interval $[t, t+\delta t]$ by a single time-independent operator.
In our calculations, we adopt a midpoint scheme for this approximation, and we verify that higher-order schemes do not affect the dynamics.

\begin{figure}[t]
    \centering \includegraphics[width=1.0\linewidth]{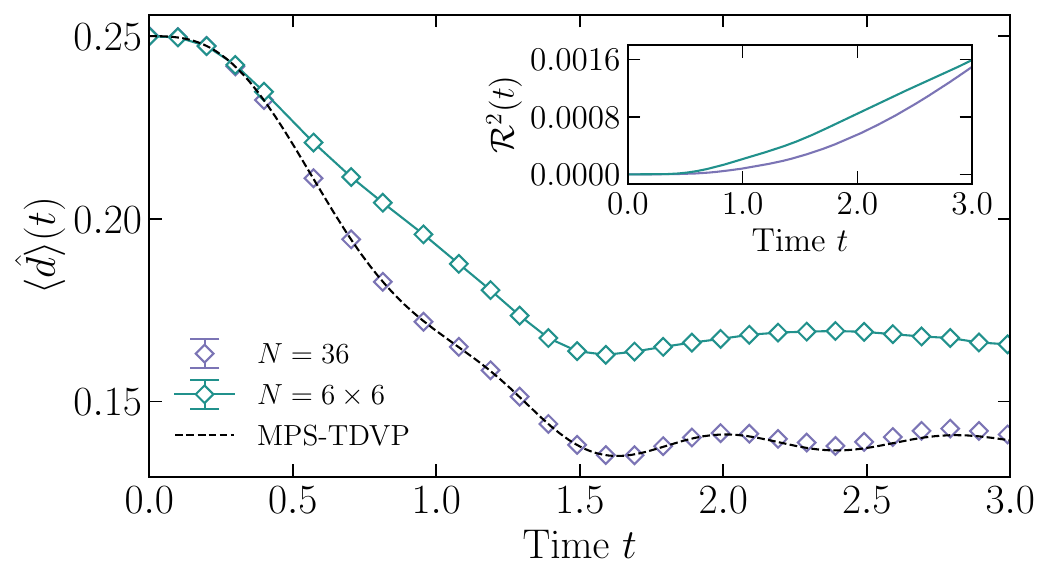}
    \caption{Time evolution of the double occupancy $\langle \hat{d} \rangle(t)$ at $U=3$ for a one-dimensional cluster with $N=36$ sites and for the $N=6 \times 6$ square lattice. tVMC calculations are indicated with empty markers.
    The dashed black line indicates the result obtained from MPS-TDVP simulation.
    The inset reports the TDVP error [see \cref{eq:integrated_error}] along the trajectories as a measure of the variational accuracy.
    The Monte Carlo errors are smaller than the markers and are not shown for clarity.}
    \label{fig:beyond_ed_benchmark}
\end{figure}

\subsubsection{Variational Error}
\label{sec:tdvp_error}

The accuracy of the variational dynamics simulated with tVMC can be quantified by the integrated TDVP error~\cite{vovrosh2025simulatingdynamicstwodimensionaltransversefield}:
\begin{equation}
\label{eq:integrated_error}
    \mathcal{R}^2 (t) = \frac{1}{\sqrt{N}} \int_0^t \sqrt{\delta s^2}, 
\end{equation}
where the residual $\delta s^2$ corresponds to: 
\begin{equation}
\delta s^2 = \delta t^2 [\text{Var}(\hat{H}) + \dot{\theta}^T S \theta - 2 \text{Re}(F)^T \dot{\theta}], 
\end{equation}
with $\text{Var}(\hat{H}) = \langle \hat{H}^2 \rangle - \langle \hat{H} \rangle^2$ denoting the variance of the possibly time-dependent Hamiltonian. 
The prefactor $1/\sqrt{N}$ in~\cref{eq:integrated_error} compensates for the linear scaling of $\text{Var}(\hat{H})$ with the system size~\cite{wu2024variational}, thereby rendering the TDVP error $\mathcal{R}^2 (t)$ size-independent and enabling meaningful comparisons between different system sizes.
We note that alternative definitions of the TDVP error have been proposed in the literature~\cite{carleo2017solving,schmitt2020quantum,nys2024ab}. 
In the majority of these approaches, $\delta s^2$ is typically normalized by $\text{Var}(\hat{H})$ directly. 
However, such definitions can become ill-conditioned when the variance is small or vanishes.
We additionally remark that $\mathcal{R}^2(t)$ is evaluated on a sample set distinct from that used to solve the tVMC equations in order to avoid underestimating the variational error due to overfitting to a specific batch of samples~\cite{hofmann2022role}.

We finally note that the integrated error in~\cref{eq:integrated_error} corresponds to the global loss function introduced in Ref.~\cite{sinibaldi2024time}, since the residual $\delta s^2$ coincides with the time-step loss function therein up to a factor of $\delta t^2$.

\subsubsection{Transformer Backflow Wave Function}\label{app:architecture}
The variational wave function employed in this work is based on a transformer neural-network architecture~\cite{vaswani2023}, originally introduced for sequence modeling and subsequently adapted to represent many-body wave functions~\cite{viteritti2023transformer_1d,viteritti2023transformer_2d,rende2025}.
The network takes as input a many-electron configuration $\boldsymbol{n}=(\boldsymbol{n}_{\uparrow},\boldsymbol{n}_{\downarrow})$, where ${\boldsymbol{n}_{\sigma}=(n_{1\sigma},\dots,n_{N\sigma})}$ for $\sigma\in\{\uparrow,\downarrow\}$, with $n_{i\sigma}\in\{0,1\}$.
Each local occupation variable $n_{i\sigma}$ is mapped to an embedding vector $\boldsymbol{x}_{i\sigma}\in\mathbb{R}^{D}$, where $D$ denotes the embedding dimension. The resulting input sequence is then given by $(\boldsymbol{x}_1,\dots,\boldsymbol{x}_N)$, with $\boldsymbol{x}_i=\boldsymbol{x}_{i\uparrow}+\boldsymbol{x}_{i\downarrow}$. Alternative embedding strategies based on tokenization of the input configuration have also been explored and yield comparable results~\cite{gu2025,rende2026alm}.

The transformer processes the input sequence and produces a new set of vectors $(\boldsymbol{y}_1,\dots,\boldsymbol{y}_N)$, with $\boldsymbol{y}_i\in\mathbb{R}^{D}$, by means of a factored attention mechanism~\cite{rende2024prr,viteritti2023transformer_1d, rende2025queries} supplemented with a spatial inductive bias~\cite{viteritti2026approaching}.
In the present work, the network architecture consists of $n_l=2$ layers, $h=12$ attention heads, and an embedding dimension $D=72$; additional details on the role of these hyperparameters can be found in Ref.~\cite{viteritti2023transformer_2d}.

The output vectors generated by the transformer are then used to define a set of single-particle backflow orbitals~\cite{diluo2019} through a site-resolved linear map~\cite{gu2025,rende2026alm},
\begin{equation}
\Phi_{i\sigma\alpha}(\boldsymbol{n})=\sum_{\beta=1}^{D} y_{i\beta}(\boldsymbol{n})\, W_{i\sigma\alpha\beta},
\end{equation}
where $W_{i\sigma\alpha\beta}$ are trainable parameters and $\alpha=1,\dots,N_e$ labels the orbitals, with $N_e$ being the total number of electrons.
By introducing the composite index $r=(i,\sigma)$, the tensor can be reshaped as $\Phi_{r\alpha}\in\mathbb{R}^{2N\times N_e}$. The wave-function amplitude is then obtained by selecting the rows corresponding to the occupied entries of the configuration $\boldsymbol{n}$ and computing the determinant
\begin{equation}\label{eq:var_wf}
\Psi_{\theta}(\boldsymbol{n})=\det\!\left[\boldsymbol{n}\star \Phi(\boldsymbol{n})\right],
\end{equation}
where $\boldsymbol{n}\star \Phi(\boldsymbol{n})$ denotes the resulting $N_e\times N_e$ matrix; further details can be found in Ref.~\cite{rende2026alm, viteritti2026beyond}.

In order to improve the variational accuracy, translational symmetry is enforced \textit{a posteriori} on the wave function in \cref{eq:var_wf}, following the symmetry-restoration scheme described in Refs.~\cite{robledo2022,sharma2025}.
A detailed analysis of the impact of symmetry restoration on the variational wave function in \cref{eq:var_wf} is provided in Ref.~\cite{viteritti2026beyond}.

\subsubsection{Validation with Other Methods}
\label{app:validation}

We first assess the accuracy of our simulations by benchmarking against exact diagonalization (ED) results whenever such calculations are feasible, namely for the $4 \times 4$ lattice.
As shown in~\cref{fig:4x4_benchmark}, our results agree excellently with ED across several values of the interaction strength $U$, supporting the accuracy of our tVMC calculations.

After this benchmarking, we further assess the accuracy of the method for system sizes beyond the reach of ED, focusing on the $6 \times 6$ lattice. 
To this end, we consider a one-dimensional system with the same number of sites, $N=36$, for which accurate MPS simulations can be performed using the TDVP algorithm~\cite{mps-tdvp}.
Therefore, we can compare the variational accuracy of the one-dimensional simulation, where a reliable reference calculation is available, with that of the $6 \times 6$ lattice, for which no direct benchmark is accessible. 
\cref{fig:beyond_ed_benchmark} shows that, for the one-dimensional cluster, the tVMC simulations are in excellent agreement with the MPS-TDVP results.
Moreover, the inset indicates that the error for the corresponding two-dimensional system remains very close to that observed in the one-dimensional case, supporting the accuracy of the simulation on this larger square lattice.

\begin{figure*}[hbtp]
    \centering \includegraphics[width=1.0\linewidth]{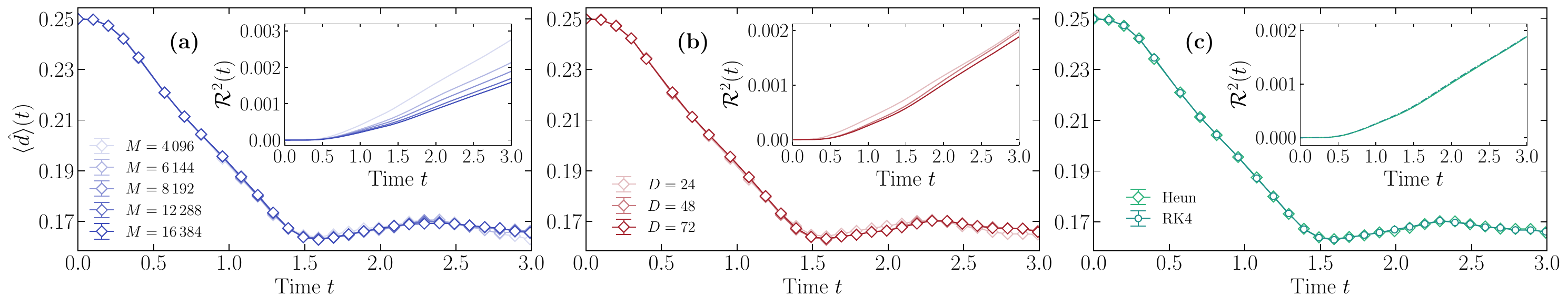}
    \caption{Time evolution with tVMC of the double occupancy $\langle \hat{d} \rangle(t)$ for different numbers of samples $M$ in the stochastic estimations [see panel \textbf{(a)}], different numbers of variational parameters $P$ [see panel \textbf{(b)}], controlled by the transformer embedding dimension $D$, and two ODE integrators [see panel \textbf{(c)}].
    The system is a $6 \times 6$ lattice with interaction strength $U = 3$.
    The inset shows the TDVP error for the trajectories as a measure of the variational accuracy.
    In panel \textbf{(a)}, we fix $D = 72$, while in panel \textbf{(b)} we fix the number of samples to $M = 8\,192$. In panel \textbf{(c)}, we use $D = 72$ and $M = 8\,192$.
    The Monte Carlo errors are smaller than the markers and are not shown for clarity.}
    \label{fig:scaling_Ns}
\end{figure*}

\subsubsection{Convergence with the Hyperparameters}
\label{app:hyperparams}

We analyzed the convergence of the tVMC simulations with respect to the key hyperparameters and algorithmic choices.
First of all, we consider the number of Monte Carlo samples and the number of variational parameters in the architecture, the latter being controlled by the transformer hidden dimension $D$.
As displayed in~\cref{fig:scaling_Ns}(a) and \cref{fig:scaling_Ns}(b), the time evolution of the observable progressively converges as both the sample size and the number of parameters are increased.
Consistently, the TDVP error decreases monotonically with both hyperparameters.
Since we observe that the results are fully converged for $M = 16\,384$ samples and hidden dimension $D = 72$, we use this setup in all subsequent calculations.
We also tested the blurred-sampling technique proposed in Ref.~\cite{wan2026removing} to mitigate the large-variance problem affecting stochastic VMC estimators~\cite{sinibaldi2023unbiasing}.
For the time evolutions considered here, however, this procedure does not lead to appreciable changes in the results.

Finally, we compared two ODE integrators for the time propagation of the tVMC equations: the second-order Heun scheme and the fourth-order Runge–Kutta method. \cref{fig:scaling_Ns}(c) shows that both the observable dynamics and the TDVP error are essentially indistinguishable for the two integrators.
We therefore employ the Heun scheme throughout, as it provides the same level of accuracy at a lower computational cost.

\subsection{Infinite-Time Dynamics}
\label{app:infinite_time}

\subsubsection{Time-Dependent Linear Variational Method}
\label{app:lvm}
To extract the infinite-time limit of the observable, we resort to the approach introduced in Ref.~\cite{sinibaldi2024time} based on the time-dependent Linear Variational Method (LVM). 
The LVM provides the equations of motion for the unitary time evolution generated by the Schrödinger equation projected onto the subspace spanned by a set of basis states.
In practice, for a linear Ansatz of the form 
\begin{equation}
    \ket{\Psi_{\alpha}(t)} = \sum_{k=1}^M \alpha_k(t) \ket{\Phi_k}
\end{equation}
with $M$ fixed normalized basis states $\ket{\Phi_k}$ and time-dependent variational coefficients $\alpha_k(t)$, the optimal trajectory of the $\alpha_k(t)$ satisfying~\cref{eq:tdse} is obtained by solving the following system of ODEs:
\begin{equation}
\label{eq:exact_coefficients}
\mathbb{S} \dot{\alpha}(t) = -i \mathbb{H} \alpha(t),
\end{equation}
where $\mathbb{S}_{kl} = \bra{\Phi_k} \ket{\Phi_l}$ is the overlap (Gram) matrix of the basis states, and $\mathbb{H}_{kl} = \bra{\Phi_k} \hat{H} \ket{\Phi_l}$ is the Hamiltonian matrix in the basis subspace~\cite{motta2024subspace}.
The system~\cref{eq:exact_coefficients} remains valid for a time-dependent Hamiltonian $\hat{H}(t)$ and can be solved by inverting $\mathbb{S}$ and integrating with any numerical scheme. 
When $\hat{H}$ does not depend on time, the solution is analytical and can be written as $\alpha(t) = \exp(-i t \mathbb{S}^{-1} \mathbb{H}) \alpha(0)$, with $\alpha(0)$ setting the initial condition.
The matrix elements of $\mathbb{S}$ and $\mathbb{H}$ can be efficiently estimated via Monte Carlo sampling from the basis states~\cite{sinibaldi2024time,entwistle2023electronic}, namely:
\begin{equation}
    \mathbb{S}_{kl} = e^{i (\text{arg} s_{kl} + \text{arg} s^*_{lk})/2}\sqrt{|s_{kl} s_{lk}^*|}, 
\end{equation}
where $s_{kl} = \mathbb{E}_{\boldsymbol{n} \sim |\Phi_k(\boldsymbol{n})|^2}[\phi_l(\boldsymbol{n}) / \phi_k(\boldsymbol{n})]$, with $\phi_k(\boldsymbol{n})$ the unnormalized amplitude of the normalized basis wave function $\Phi_k(\boldsymbol{n})$.
An equivalent expression including $\hat{H}$ is valid for $\mathbb{H}_{kl}$.
In all calculations, we used $M =524\,288$ samples per basis state, as we observed convergence in both the entries and the eigenspectrum of the matrices with this sample size.
We remark that~\cref{eq:exact_coefficients} can alternatively be solved using a multidimensional extension of VMC, which employs a determinant constructed from the basis states and thereby avoids the explicit inversion of $\mathbb{S}$~\cite{pfau2024accurate,kahn2026variational,hendry2025grassmannvariationalmontecarlo}.
However, there is no general procedure for extending the Monte Carlo transition rule to this multidimensional setting, making sampling a challenging aspect.
Furthermore, in our settings, we did not encounter the linear-dependence instabilities reported in Ref.~\cite{kahn2026variational} that this determinant-based formulation aims to address. 
We therefore adopted the standard method outlined above, foregoing the more involved multidimensional approach.

\begin{figure*}[t]
    \centering \includegraphics[width=1.0\linewidth]{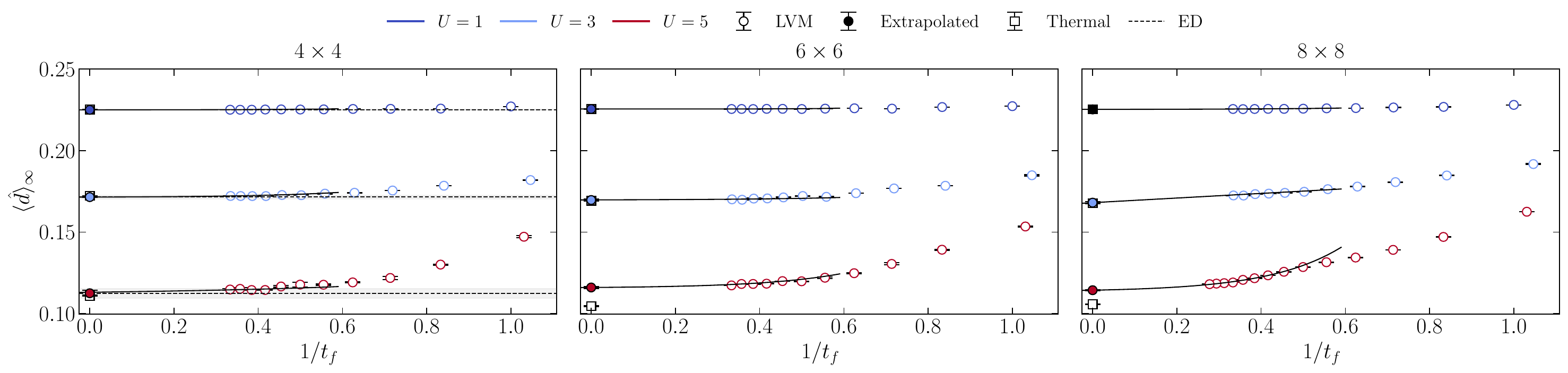}
    \caption{
    Double occupancy at infinite time $\langle \hat{d} \rangle_{\infty}$ for various interaction strengths $U$ and system sizes $4\times 4$, $6\times 6$ and $8 \times 8$, computed using the time-dependent LVM, as a function of the final time $t_f$ of the tVMC simulation. Empty markers show the LVM data. 
    The last data are fitted with the functional form $a + b(e^{x/c} - 1)$ represented by solid lines, and filled markers indicate the extrapolated values in the limit $t_f \rightarrow \infty$. 
    The statistical error on $\langle \hat{d} \rangle_{\infty}$ is estimated by repeating the LVM calculation over independent sets of Monte Carlo samples, and propagated to the extrapolated values via a Gaussian resampling procedure. 
    The thermal expectation values are shown at $1/t_f = 0$ for comparison. For the $4 \times 4$ system, the exact diagonalization  results are included as dashed horizontal lines, obtained by averaging the trajectory over a sufficiently long late-time window. The shaded grey bands represent the uncertainties of the ED results, estimated using block averaging.
    }
    \label{fig:extrapolation}
\end{figure*}

The LVM framework provides a rigorous method to extract the infinite-time limit of any observable, thanks to the linearity of the Ansatz $\ket{\Psi_{\alpha}(t)}$.
By expanding the trajectory of the coefficients in a Fourier series $\alpha_k(t) = \sum_l \gamma_{kl} e^{i \omega_l t}$, and dropping all oscillating terms in the expectation value of an observable $\hat{A}$, we obtain the closed expression:
\begin{equation}
\label{eq:thermalization}
\frac{\bra{\Psi_{\alpha}(t)} \hat{A} \ket{\Psi_{\alpha}(t)}}{\bra{\Psi_{\alpha}(t)} \ket{\Psi_{\alpha}(t)}} \overset{t \rightarrow \infty}{\approx}  \frac{\sum_{klm}\gamma^*_{km} \mathbb{A}_{kl} \gamma_{lm}}{\sum_{klm}\gamma^*_{km} \mathbb{S}_{kl} \gamma_{lm}}, 
\end{equation}
where $\mathbb{A}_{kl} = \bra{\Phi_k} \hat{A} \ket{\Phi_l}$. 
Other strategies extend the dynamics to long times by directly extrapolating the time series of observables~\cite{imadalongtimedynamics,gulllongtimedynamics}.
The LVM approach differs fundamentally in that it extrapolates the quantum state itself, making it inherently a more \emph{ab initio} method.
For the application considered here, we find that this leads to remarkably better results.

The accuracy of the LVM method depends on how well the subspace spanned by the basis states captures the actual subspace spanned by the quantum dynamics.
As basis states, we use tVMC snapshots equally spaced by a time interval $\Delta t$.
This choice is well-motivated: in the limit where the snapshots coincide with the exact 
propagated states and are sampled sufficiently densely in time, the LVM dynamics reduce to 
the exact time evolution.
Moreover, equally spaced snapshots naturally form a Krylov basis of the unitary time-step propagator, $\ket{\Phi_k}  = (e^{-i \Delta t \hat{H}})^k \ket{\Psi_0}$, which is known to provide a representative basis for 
subspace methods~\cite{motta2024subspace,kahn2026variational}.

Assuming the basis states are sufficiently representative, their number $M$ is a key control parameter for the accuracy of the LVM simulation.
For equally spaced snapshots separated by $\Delta t$, $M$ is proportional to the total tVMC simulation time $t_{f}$ via $t_f = M \Delta t$. 
Therefore, we can study the convergence of the LVM calculation as a function of $t_f$ and safely extrapolate the limit of $t_f \rightarrow \infty$.
In our calculations, we consider $\Delta t = 0.2$.

In~\cref{fig:extrapolation}, we show the infinite-time double occupancy, $\langle \hat{d} \rangle_{\infty}$, computed with LVM for different system sizes and interaction strengths $U$, as a function of the final tVMC time $t_f$.
The data are fitted with an exponential function, which accurately captures the observed behavior, and the limit $t_f \to \infty$ is then obtained by extrapolating the fit.
For the $4 \times 4$ system, the extrapolated value agrees well with $\langle \hat{d} \rangle_{\infty} $ obtained from the time evolution computed with exact diagonalization, supporting the reliability of the extrapolation procedure. We observe that the dependence on $t_f$ becomes more pronounced as $U$ increases, indicating that longer tVMC trajectories are required to obtain a stable extrapolation at stronger interactions.
This behavior is expected: in the trivial limit $U=0$, the initial state is already an eigenstate of the Hamiltonian and a single LVM basis state is sufficient to reproduce the exact dynamics.
For increasing $U$, the dynamics explores a progressively larger subspace, so that longer tVMC trajectories, and hence more LVM basis states, are needed.
The extrapolation $t_f \to \infty$ is designed to account for this finite-basis effect, reducing the residual error associated with using a finite set of variational states.
The fit remains robust for the larger system sizes, where we find that the extrapolated value is consistent with the thermal average for $U=1$ and $U=3$, whereas a significant deviation is observed for $U=5$, suggesting a substantial delay of thermalization at stronger interaction strengths.

\subsection{Canonical Finite-Temperature Calculation}
\label{app:finite_temperature}
The thermal expectation values are computed using AFQMC~\cite{blankenbecler1981monte}. 
To ensure a meaningful comparison with the dynamics, the simulations are performed in the canonical ensemble at a fixed particle number with zero $z$-magnetization, since these quantum numbers are fixed by the initial state and conserved by the time evolution.
In practice, we enforce these constraints following Ref.~\cite{wang2017finite}, in which grand-canonical AFQMC simulations are performed with an augmented Hamiltonian that exponentially suppresses deviations in particle number and magnetization:
\begin{equation}
    \hat{H}_{\text{can}} = \hat{H}(t=\tau) + \lambda_N (\hat{N}_P - N_P)^2 + \lambda_S(\hat{S}^{z} - S^z)^2,
\end{equation}
where $H(t)$ is from~\cref{eq:hamiltonian}, $\hat{N}_P = \sum_i (\hat{n}_{i\uparrow} + \hat{n}_{i\downarrow})$ is the total particle number 
operator and $\hat{S}^z = \sum_i (\hat{n}_{i\uparrow} - \hat{n}_{i\downarrow})/2$ is the total spin operator along the $z$ axis.
The scalars $N_P$ and $S^z$ correspond to the target number of particles and magnetization, and in our half filling case, they are set to $N_P=N$ and $S^z=0$.
The parameters $\lambda_N$ and $\lambda_S$ control the penalty strengths and must be large enough to enforce the constraints.
Practically, we  compute the fluctuations of  the total particle number and magnetization, and enhance the values of  $\lambda_N$ and $\lambda_S$ until these fluctuations are compatible with zero within error bars. 
We remark that the projection onto the $S^z=0$ subspace generates a mild sign problem.
As an example, for the $L=8$ data point at $U/t = 4$, the average sign is $\overline{s}  = 0.929 \pm 0.001$. However, since $\Delta \overline{s} / \overline{s}  \ll 1 $, division by $\overline{s}$ poses no numerical problem. 
To determine $T_{\text{eff}}$ from~\cref{eq:t_eff}, we compute the thermal energy for several values of $T_{\text{eff}}$, fit the resulting energy curve with a polynomial function, and invert this fit to extract the value of $T_{\mathrm{eff}}$ that matches the energy obtained from the dynamics.

\section{Data availability}
The numerical simulations are performed using NetKet~\cite{netket2,vicentini2022netket}, with the code to be made publicly available in a future revision. Exact diagonalization is carried out with XDiag~\cite{xdiag}, while Matrix Product State simulations are conducted with ITensor~\cite{fishman2022itensor}. 
The finite-temperature Auxiliary-Field Quantum Monte Carlo calculations are realized with the ALF library~\cite{alf}, and the augmented Hamiltonian for performing the canonical averages can be found on GitHub~\cite{hubbardcanonical}.

\begin{acknowledgments}
We thank F. Becca, M. Imada, Y. Nomura, Y. Murakami, R. Rossi, L. Mauron, J. Nys, M. Eckstein, P. Werner, and M. Heyl for fruitful discussions.
This work was supported as part of the ``Swiss AI initiative'' by a grant from the Swiss National Supercomputing Centre (CSCS) under project ID a117 on Alps.
AS is supported by the Google PhD Fellowship 2025. 
RR and LLV acknowledge the CINECA award under the ISCRA initiative for the availability of high-performance computing resources and support.
RR acknowledges support from the Flatiron Institute.
The Flatiron Institute is a division of the Simons Foundation.
LLV is supported by SEFRI under Grant No. MB22.00051 (NEQS - Neural Quantum).
FFA acknowledges scientific support and HPC resources provided by the Erlangen National High Performance Computing Center (NHR@FAU) of the Friedrich-Alexander-Universit\"at Erlangen-N\"urnberg (FAU) under the NHR project b133ae. NHR funding is provided by federal and Bavarian state authorities. NHR@FAU hardware is partially funded by the German Research Foundation (DFG) -- 440719683. FFA acknowledges support by the W\"urzburg-Dresden Cluster of Excellence {\it ctd.qmat} (EXC 2147, Project No.~390858490).
\end{acknowledgments}

\bibliography{biblio}

\end{document}